\documentclass{pas}

\usepackage{multirow}
\usepackage{newtxtext}
\usepackage{newtxmath}
\usepackage{tabularx}
\usepackage[T1]{fontenc}

\usepackage{graphicx}	
\usepackage{amsmath}	
\usepackage{longtable}
\usepackage{booktabs}
\usepackage[table]{xcolor}
\RequirePackage{etex}

\begin{document}

\lefttitle{Thirty years of primordial deuterium abundances}
\righttitle{Richard de Grijs and Giuseppe Bono}

\jnlPage{1}{18}
\jnlDoiYr{2026}
\doival{10.1017/pasa.xxxx.xx}

\articletitt{Research Paper}

\title{Thirty years of primordial deuterium abundances: A provenance-aware statistical reanalysis}

\author{\sn{Richard} \gn{de Grijs}$^{1,2}$ and \sn{Giuseppe} \gn{Bono}$^{3,4}$}

\affil{$^{1}$School of Mathematical and Physical Sciences, Macquarie University, Balaclava Road, Sydney, NSW 2109, Australia\\
$^{2}$Astrophysics and Space Technologies Research Centre, Macquarie University, Balaclava Road, Sydney, NSW 2109, Australia\\
$^{3}$Dipartimento di Fisica, Universit\`a di Roma Tor Vergata, via Della Ricerca Scientifica 1, I-00133, Rome, Italy\\
$^{4}$INAF, Rome Astronomical Observatory, via Frascati 33, I-00078 Monte Porzio Catone, Italy}

\corresp{Richard de Grijs, Email: richard.de-grijs@mq.edu.au}

\citeauth{Richard de Grijs and Giuseppe Bono, Thirty years of primordial deuterium abundances: A provenance-aware statistical reanalysis. {\it Publications of the Astronomical Society of Australia} {\bf 00}, 1--18 (2026). https://doi.org/10.1017/pasa.xxxx.xx}

\history{(Received xx xx xxxx; revised xx xx xxxx; accepted xx xx xxxx)}

\begin{abstract}
Primordial deuterium provides one of the most precise observational tests of Big Bang nucleosynthesis, yet the published literature is built on a comparatively small number of high-redshift absorption systems that are repeatedly observed, re-measured, reanalysed and incorporated into compilations. We construct a provenance-aware database of 957 published primordial-D/H values from 247 publications spanning 1994--2026 and trace observational entries, wherever possible, to their underlying physical absorbers. Of the 795 observational entries, 80.1 percent represent reused individual or compilation values; only 11.4 percent correspond to new determinations of individual absorption systems. Although 52 distinct absorbers are identified overall, the eight most frequently represented systems account for 62.0 percent of all identifiable absorber mentions. An inverse Herfindahl--Hirschman measure applied to our full database gives an effective representation equivalent to only 16.28 equally represented absorbers, demonstrating that the published literature is strongly concentrated around a much smaller effective subset. Despite this strong internal dependence, the modern consensus value of $10^5({\rm D/H})\simeq2.5$ is robust to alternative absorber-level treatments and survives removal of the eight dominant systems. However, the apparent precision of literature-level compilations is more sensitive to repeated representation than a simple count of published values implies. We also identify a synchronised transition in the compilation literature around 2013--2014, associated with increasingly selective and homogeneous analyses while older individual values continued to circulate. The modern D/H consensus is therefore robust in central value, but the underlying literature is less statistically independent than its publication density suggests, because many reported values share the same absorbers, measurements or compilation inputs. This thus highlights the importance of provenance-aware analysis in precision cosmology.
\end{abstract}


\maketitle

\section{Introduction}

Primordial deuterium occupies a distinctive position among the light-element abundances produced during Big Bang nucleosynthesis (BBN). Its abundance is highly sensitive to the cosmic baryon density, a long-established result of standard BBN calculations (e.g. \citealt{1967ApJ...148....3W,1984ApJ...281..493Y,1996Natur.381..207T}; see also \citealt{2025qal1.book.....C}). Subsequent stellar processing almost exclusively destroys rather than produces deuterium. Measurements of the deuterium-to-hydrogen abundance ratio, D/H, in chemically primitive environments therefore provide one of the most direct observational links between the baryonic content of the early Universe and the predictions of standard BBN. In combination with increasingly precise determinations of the baryon density from the cosmic microwave background, the primordial deuterium abundance has consequently become an important test of the internal consistency of the standard cosmological model.

During the past three decades, the observational determination of primordial D/H has undergone a striking transformation. The first measurements in high-redshift (damped Lyman-$\alpha$) quasar absorption systems exhibited large scatter, substantial uncertainties and, in several cases, values differing by more than an order of magnitude \citep{1994MNRAS.268L...1C, 1994Natur.368..599S, 1994AAS...185.7004T, 1996MNRAS.278..506C, 1996PhRvD..54.3681K, 1996AJ....111.2135R, 1996ApJ...459L...1R, 1996Natur.381..207T, 1996AnA...316...33W}. Improvements in spectral resolution, signal-to-noise ratio, absorption-line modelling and the identification of suitably simple low-metallicity absorption systems subsequently produced much tighter constraints. Modern determinations now cluster closely around a primordial abundance of $10^5 {\rm (D/H)} \simeq 2.5$, with individual measurements in some cases claiming percent-level precision \citep[e.g.,][]{2014ApJ...781...31C, 2016ApJ...830..148C, 2018ApJ...855..102C, 2015MNRAS.447.2925R, 2017MNRAS.468.3239R, 2016MNRAS.458.2188B, 2018MNRAS.477.5536Z, 2024MNRAS.529..839G, 2024MNRAS.528.4068K}.

This apparent precision, however, conceals an unusual feature of the underlying literature. Unlike many cosmological observables for which increasingly large surveys provide progressively larger statistically independent samples, primordial deuterium measurements remain based on a comparatively small number of suitable absorption systems. Individual absorbers are repeatedly reanalysed, incorporated into successive compilations and cited as contributors to later primordial-abundance estimates. As a result, the number of published D/H values greatly exceeds the number of physically independent systems from which the observational information originates (see also Section \ref{growth.sec}).

This distinction matters whenever the published literature is interpreted statistically. A compilation containing hundreds of reported primordial-D/H values is not equivalent to hundreds of independent measurements if many entries ultimately originate from the same spectra, absorbers or previously published numerical estimates. Moreover, repeated use of particular values can give individual measurements or compilations a persistence in the literature that is not captured by conventional inverse-variance weighting. Understanding the provenance of published values is therefore essential both for reconstructing the historical emergence of the modern consensus and for assessing its effective statistical independence (see Section~\ref{robust.sec}).

Over the past decade, we have used comprehensive literature compilations to examine the temporal evolution, dispersion and apparent convergence of published astrophysical measurements, initially in a series of studies of Galactic and extragalactic distance and dynamical estimates \citep{2014AJ....147..122D, 2014AJ....148...17D, 2015AJ....149..179D, 2016ApJS..227....5D, 2017ApJS..232...22D, 2020ApJS..246....3D, 2020ApJS..248....6D} and, more recently, in a statistical reanalysis of published primordial-helium abundances \citep{2026PASA...43...82H}. These studies showed that increasingly tight agreement among published values can reflect a combination of genuine methodological improvement, shared data and/or analysis frameworks, and the repeated use of the same influential measurements or benchmarks. The present study extends this approach to primordial deuterium, where the problem of statistical dependence is particularly acute, because a large published literature ultimately rests on a comparatively small number of high-redshift absorption systems.

Building on our earlier analyses of publication patterns in extragalactic distance measurements and primordial helium abundances, we have assembled a comprehensive database of published observational determinations and BBN predictions published from 1994 onwards, retaining not only central values and uncertainties but also the identities of the underlying absorption systems and, wherever possible, the provenance of reused measurements and compilations. Our aim is not to derive a new preferred value of primordial D/H. Instead, we ask how the modern consensus emerged, how much genuinely new observational information entered the literature over time, how strongly that information is concentrated among a small number of absorbers and how previously published values subsequently propagated through the literature.

The remainder of this paper is organised as follows. Section~\ref{data.sec} describes the construction, classification and provenance structure of the database. Section~\ref{structure.sec} reconstructs the historical development of the primordial-D/H literature, including the growth and concentration of the underlying absorber sample and the numerical evolution of the dominant systems. Section~\ref{statistics.sec} quantifies the statistical consequences of repeated measurements, absorber-level clustering and numerical reuse. In Section~\ref{discussion.sec} we interpret these results in terms of the emergence of the modern consensus, compilation-level inheritance and the implications for precision cosmology. Our conclusions are summarised in Section~\ref{conclusions.sec}.

\section{Data}
\label{data.sec}

\subsection{Literature search and compilation}

We adopt 1994 as the starting point of the present compilation because that year marks the emergence of direct measurements of deuterium in high-redshift quasar absorption systems. \citet{1994Natur.368..599S} reported a probable detection of deuterium in the $z=3.32$ absorber towards Q0014+813, thereby initiating the observational programme that eventually developed into precision measurements of the primordial deuterium abundance. The identification was subsequently regarded with increasing caution because unrelated H{\sc i} absorption can mimic the expected D{\sc i} feature, and this early high-D/H system does not form part of the modern high-precision primordial-deuterium samples. Earlier studies had established the cosmological significance of deuterium \citep[e.g.,]{1976AnA....50..461A} and had measured D/H in Galactic, Solar System and other astrophysical environments \citep[e.g.,][]{1972AnA....18..126G, 1973ApJ...186L..95R}, but they do not form part of the high-redshift absorption-system literature considered here.

Candidate publications were identified through an extensive search of the NASA Astrophysics Data System (ADS). The initial ADS search was deliberately broad and was intended to maximise recall rather than precision. We inspected all 4931 records published between 01/1994 and 05/2026 for which `deuterium' appeared in the abstract. The initial search was deliberately broad so as not to impose assumptions about how primordial-D/H results were described by their authors. Titles and abstracts were screened manually, and papers concerned exclusively with molecular deuterium, Solar System abundances, stellar or interstellar-medium measurements, and other applications unrelated to primordial nucleosynthesis were excluded. Potentially relevant papers were then examined individually, including their tables, figures and contextual discussion where necessary. We also followed citation and reference trails when these revealed additional published primordial-D/H values not identified directly by the initial search.

The resulting compilation was designed to capture papers presenting new measurements of primordial D/H as well as subsequent reanalyses, compilations and theoretical BBN predictions that explicitly reported numerical values of D/H. This broader scope is important for the present study. Our aim is not simply to reconstruct the sequence of independent observations but to characterise how numerical estimates of primordial D/H entered, evolved and then propagated through the published literature.

The final database contains 957 reported primordial-D/H values drawn from 247 distinct publications.\footnote{An electronic version of the database is available at \url{http://astro-expat.info/Data/pubbias.html}. A permanent link to this page can be found at \url{http://web.archive.org/web/20260812084630/https://astro-expat.info/Data/pubbias.html}; members of the community are encouraged to send us updates or missing information.} In addition, 113 references (also included in our electronic database file) were recorded separately to document the provenance of reused values, absorber identifications and sample relationships. These provenance references do not necessarily correspond to papers from which D/H values were extracted directly; rather, they support the reconstruction of how published measurements and compilations relate to earlier work.

Because no independent complete census of the primordial-deuterium literature exists, we cannot assign a formal completeness or purity statistic to this selection procedure. The search strategy was intentionally high-recall and was supplemented by manual reference tracing; conversely, inclusion in the final database required the presence of a numerical primordial-D/H measurement, compilation value or BBN prediction relevant to the scope of this study. We therefore regard the resulting database as a comprehensive literature compilation, while recognising that no manually assembled bibliographic sample can be guaranteed to be absolutely complete.

As in our earlier studies of distance measurements and our recent statistical reanalysis of the primordial helium abundance, we treat the published literature itself as an object of study. The emphasis is consequently on completeness, temporal evolution and statistical dependence rather than on selecting a restricted subset of measurements judged \textit{a priori} to be the most reliable.

\subsection{Extraction of published D/H values}

For each relevant publication we recorded every numerical estimate of the primordial D/H abundance ratio that could be associated with an observational determination, an observational compilation or a BBN prediction. Values were homogenised to the conventional form, $10^5({\rm D/H})$. Where asymmetric uncertainties were reported, the upper and lower uncertainties were retained separately. Parameter ranges, as well as upper and lower limits, were likewise retained.

For each database entry we recorded, where available,

\begin{itemize}
    \item the bibliographic identifier and publication date;
    \item the reported value of $10^5({\rm D/H})$ or equivalent (e.g., if rendered in logarithmic form);
    \item the quoted upper and lower uncertainties;
    \item the type of determination or prediction;
    \item the absorber name(s) as given in the source publication;
    \item canonical absorber designation(s);
    \item the absorber redshift(s);
    \item information about the observational or bibliographic provenance of the value;
    \item notes required to resolve ambiguous cases or relationships with earlier work.
\end{itemize}

Publication dates were recorded to month precision where this information was available. For publications for which only the year could be established in the NASA/ADS database, the month was recorded as zero and the corresponding decimal publication date was assigned to the integer year itself (e.g., 1997.000).

Multiple values reported in a single publication were retained as separate database entries where they represented distinct numerical determinations, alternative analyses, different absorption systems or explicitly different theoretical predictions. This choice reflects the focus of the present study on the numerical content of the published record. It does not imply that such entries are statistically independent; their relationships are instead captured through the classification and provenance information described below.

The database also retains repeated numerical values when a later publication explicitly reuses an earlier measurement or compilation. Such entries are essential for distinguishing the production of new information from its subsequent propagation. Omitting them would yield a cleaner catalogue of measurements, but it would remove precisely the historical structure that we aim to quantify.

\subsection{Classification of published values}

Each database entry was assigned to one of six categories designed to distinguish genuinely new observational or theoretical information from subsequent reuse:

\paragraph{\texttt{OBS\_INDIV\_NEW}} A new observational determination for an individual absorption system, based on newly presented observations or a new analysis of extant observations contributing genuinely new information for that specific system.

\paragraph{\texttt{OBS\_COMP\_NEW}} A newly published observational compilation or combined sample result, incorporating two or more individual absorption systems into a new estimate of primordial D/H.

\paragraph{\texttt{OBS\_INDIV\_REV}} A repeated or otherwise reused observational value associated with an individual absorption system, where the underlying measurement had already entered the literature.

\paragraph{\texttt{OBS\_COMP\_REV}} A repeated compilation value whose underlying observational content had previously been published.

\paragraph{\texttt{BBN\_PRED\_NEW}} A newly calculated theoretical prediction of primordial D/H within a BBN framework, with further details provided in the table notes.

\paragraph{\texttt{BBN\_PRED\_REV}} A theoretical D/H prediction reproduced or reused from an earlier BBN calculation rather than representing a new calculation.

The distinction between `new' and `reused' is deliberately based on the provenance of the information and not on whether a number first appears in a particular paper. A publication may, for example, present a new scientific discussion while quoting an observational D/H value derived entirely from an earlier analysis; such a value is classified as reused. Conversely, a new compilation may contain no newly observed absorber but nevertheless constitute a new numerical estimate obtained by combining previously published measurements and is therefore classified as \texttt{OBS\_COMP\_NEW}.

This classification does not imply a hierarchy of scientific quality. Reanalysis and compilation are fundamental components of cumulative science, and a later reused value may be methodologically preferable to an earlier determination. The purpose of our classification is instead to distinguish the introduction of new information from its later reinterpretation and propagation.

Observational and theoretical values are likewise kept conceptually distinct. Although BBN predictions and high-redshift measurements increasingly converge on closely similar values of primordial D/H, they are not interchangeable measurements of the same observational dataset. We therefore do not combine them into a single statistical population except where their historical relationship is itself the subject of analysis.

\subsection{Absorber identification and provenance}

A significant complication in the primordial-deuterium literature is that the number of published values greatly exceeds the number of physical absorption systems on which those values ultimately depend. The same absorber may appear under slightly different designations in different publications, be incorporated repeatedly into later compilations and contribute more than once to a single paper when alternative determinations are discussed.

We therefore homogenised absorber names to a canonical nomenclature while retaining the original published designation separately. The canonical names were established by cross-referencing coordinates, redshifts and the naming conventions used in the source literature. Where more than one physically distinct absorber occurs along the same line of sight, these were retained as separate systems. For example, \mbox{PKS1937--101} and \mbox{PKS1937--101 (2)} denote distinct absorbers at different redshifts of $z = 3.572$ versus $z = 3.256$ rather than duplicate entries. The same convention is used, where necessary, for other sightlines.

For compilation entries containing multiple absorbers, the canonical absorber field records the complete contributing list wherever this could be reconstructed. These lists were subsequently split into their constituent systems for object-level analyses (see below). An absorber may therefore contribute once to the frequency count for each database entry in which it occurs. Repeated appearances of the same absorber within a source publication were retained where they corresponded to genuinely distinct D/H values rather than being removed automatically as duplicates.

Provenance was established at several levels. In the strongest cases, the source publication explicitly identified the individual absorber or absorbers contributing to the quoted value. In other cases the connection could be inferred from the stated sample, cited source papers or contextual discussion. Where such attribution was probable but not explicit, this was recorded as an inferred identification rather than treated as direct provenance. A small number of generic values (26 of 795 observational data points, or 3.3 percent) could not be associated securely with an individual absorber; those entries were excluded from object-level analyses while remaining part of the broader literature database.

Provenance assignments were supported by our separate bibliography of 113 references cited in the provenance field. These references were derived separately from the 247 source publications from which the database entries themselves were extracted. This provenance structure allows analysis of the published record at two different levels. At the level of individual database entries, the compilation contains 957 reported values (795 observational data points and 162 BBN-derived values). At the level of underlying observational systems, however, the information is concentrated among only a few dozen high-redshift absorbers (see Section \ref{concentration.sec}). This distinction between the number of published numerical values and the number of independent physical systems provides the basis for the historical and statistical analyses that follow.

\section{The historical structure of the primordial-D/H literature}
\label{structure.sec}

Having established both the numerical content and provenance structure of the database, we first examine how the primordial-D/H literature developed historically before turning to formal statistical analysis. The five figures in this section trace the evolution of published values, the growth of the underlying observational sample, the concentration of the literature around a small number of absorbers and the subsequent propagation and convergence of measurements associated with those systems.

\subsection{Emergence and convergence of published values}

\begin{figure}
    \centering
    \includegraphics[width=\columnwidth]{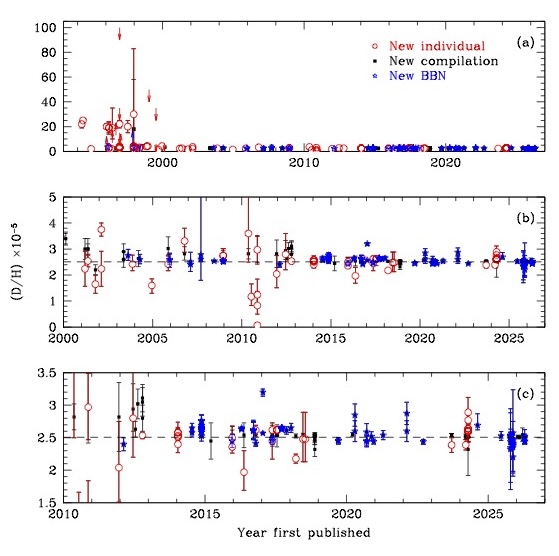}
    \caption{Evolution of newly published primordial-deuterium abundances as a function of publication date. Red open circles denote new determinations for individual absorption systems (\texttt{OBS\_INDIV\_NEW}), black filled squares represent new observational compilation values (\texttt{OBS\_COMP\_NEW}) and blue stars are new BBN predictions (\texttt{BBN\_PRED\_NEW}). Where available, quoted uncertainties are shown; arrows indicate upper or lower limits. Panel (a) shows the full 1994--2026 dataset on a linear scale to highlight the prevailing scatter, whereas panels (b) and (c) provide successive enlargements of the increasingly narrow range occupied by later determinations. The horizontal dashed line marks the PDG's 2025 recommended observational value, $10^5({\rm D/H})_{\rm p}=2.508$ \citep[][2025 update]{PhysRevD.110.030001}.}
    \label{fig1}
\end{figure}

Figure~\ref{fig1} shows the temporal evolution of newly published primordial-D/H values, distinguishing new determinations for individual absorbers, new observational compilations and new BBN predictions. The full historical range is shown in the top panel, with successive enlargements in the middle and bottom panels to emphasise the progressive contraction in central D/H values of the published distribution and the decreasing formal uncertainties. We opted to use this approach to visualisation to highlight the scatter dominating at different times across the observation period.

The earliest measurements were characterised by substantial scatter. During the mid-1990s, reported values of $10^5({\rm D/H})$ spanned more than an order of magnitude, and several results were reported as limits rather than well-constrained measurements. Much of this early dispersion reflects the observational difficulty of identifying deuterium absorption unambiguously in complex high-redshift systems, where contamination by unrelated H{\sc i} absorption can mimic the expected displacement of the deuterium lines \citep[e.g.,][]{1999AnA...349L..13M, 2000ApJ...529..655K, 2000PhR...333..409T, 2001ApJ...552..718O}. The infancy years of the field therefore represent an exploratory phase in which both the practical feasibility of the method(s) and the range of plausible primordial abundances were still being established.

By the late 1990s and early 2000s, the range of newly reported values had narrowed substantially. Individual-system determinations increasingly clustered within $2 \lesssim 10^5({\rm D/H})\lesssim4$, while large values, $10^5({\rm D/H}) \gtrsim 4$, mostly disappeared from the literature. This contraction coincided with the increasing availability of high-resolution, high-signal-to-noise echelle spectroscopy from instruments such as the High Resolution Echelle Spectrometer (HIRES) on the W.M. Keck I Telescope or, more recently, the Ultraviolet and Visual Echelle Spectrograph (UVES) on the European Southern Observatory's Very Large Telescope. 

At the same time, observational programmes became increasingly selective in identifying the relatively rare absorption systems best suited to precision D/H measurements. These include low-metallicity systems with comparatively simple and well-resolved velocity structures, such as HS0105+1619, Q1243+307, Q0913+072, J1558--0031, J1419+0829 and J1358+6522. In favourable cases, multiple H{\sc i} Lyman-series transitions constrain the neutral-hydrogen column density and component structure \citep[e.g.,][]{2012MNRAS.426.1427O, 2016MNRAS.458.2188B, 2024MNRAS.529..839G}, while corresponding D{\sc i} absorption can be distinguished from unrelated H{\sc i} interlopers \citep[e.g.,][]{2001ApJ...559...23K}. The resulting increase in precision (reflected by the narrower spread in central values and the smaller error bars) therefore reflects both improved instrumentation and increasingly stringent selection of intrinsically favourable systems (for a discussion, see also Section~\ref{inheritance.sec}).

A further transition is apparent from the early 2010s onwards. Both new observational determinations and newly calculated BBN predictions become concentrated within a narrow interval around the present-day reference abundance, $10^5({\rm D/H})_{\rm p}=2.508$ \citep[Particle Data Group (PDG) 2025:][2025 update]{PhysRevD.110.030001}. The quoted uncertainties simultaneously decrease, in some cases to the percent level. The modern literature therefore presents an appearance of striking concordance between high-redshift absorption-system measurements and BBN calculations.

This convergence should not, by itself, be interpreted as evidence of publication bias or other non-random effects. Improved spectroscopy, more sophisticated absorption-line modelling, better characterisation of systematic uncertainties and advances in nuclear-reaction calculations all naturally reduce both the scatter and uncertainties of published results. Nevertheless, Figure~\ref{fig1} establishes the central empirical feature that motivates the analyses that follow below: the primordial-D/H literature has evolved from a highly dispersed set of early measurements into a tightly clustered modern consensus.

\subsection{Growth of the literature versus growth of the observational foundation}
\label{growth.sec}

\begin{figure}
    \centering
    \includegraphics[width=\columnwidth]{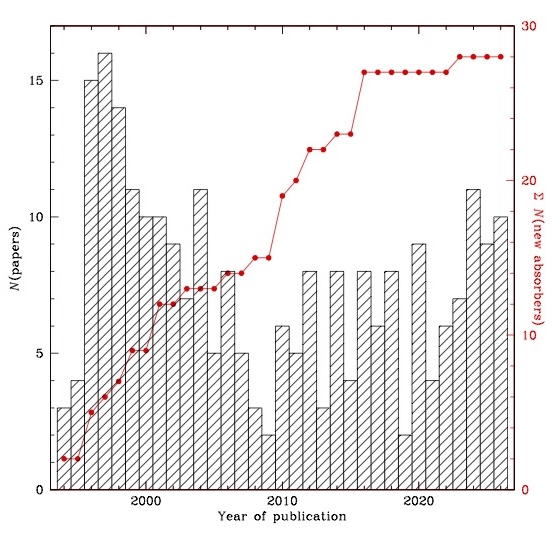}
    \caption{Growth of the published primordial-D/H literature compared with the growth of its explicitly identifiable new observational foundation. The hatched histogram shows the number of distinct publications containing D/H values in each calendar year (left-hand axis). Red points and the connecting line show the cumulative number of absorption systems entering our database through genuinely new individual determinations classified as \texttt{OBS\_INDIV\_NEW} (right-hand axis).}
    \label{fig2}
\end{figure}

The growth of the published literature does not correspond to a comparable growth in the number of independent physical systems from which primordial D/H is measured. Figure~\ref{fig2} compares the number of source publications per calendar year (left-hand $y$ axis) with the cumulative number of absorption systems contributing genuinely new individual measurements (right-hand $y$ axis).

Across the full 1994--2026 interval, the database contains D/H values extracted from 247 distinct publications. Figure~\ref{fig2} tracks the cumulative number of absorption systems that enter the database through genuinely new individual observational determinations, i.e., entries classified as \texttt{OBS\_INDIV\_NEW}. Using this definition, only 28 absorbers contribute to the cumulative curve. This number should not be confused with the 52 distinct physical absorbers identified across the database as a whole. The latter total also includes systems that appear only in compilations, reused measurements or provenance reconstructions, without a corresponding \texttt{OBS\_INDIV\_NEW} entry in our compilation. Figure~\ref{fig2} therefore measures the growth of the explicitly identifiable stream of new individual measurements rather than providing a census of every absorber represented anywhere in the literature.

This divergence in source publications versus new absorbers is important. It demonstrates that the expansion of the literature has increasingly reflected the continued analysis and reuse of an established observational foundation rather than the discovery of large numbers of new suitable absorbers. A count of published D/H values or papers therefore substantially overstates the number of statistically independent observational constraints.

The contrast is especially relevant for attempts to combine literature measurements. Treating each published value as an independent data point would effectively give repeatedly analysed absorbers greater statistical weight simply because they appear more often in the literature. The provenance information assembled here allows this distinction between publication frequency and observational independence to be made explicitly.

\subsection{Concentration of the literature among individual absorbers}
\label{concentration.sec}

\begin{figure}
    \centering
    \includegraphics[width=\columnwidth]{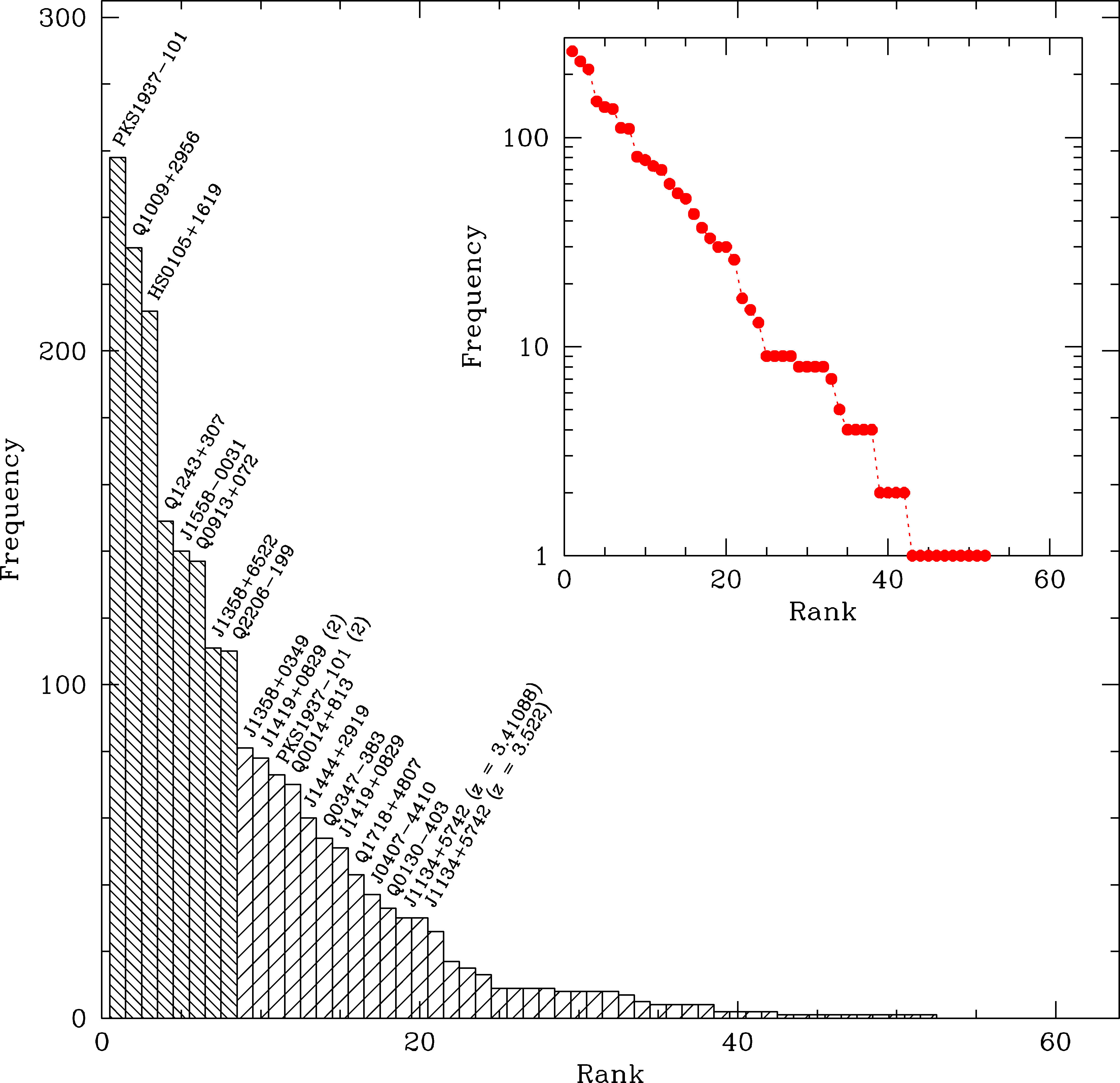}
    \caption{Ranked frequency distribution of the 52 distinct physical absorption systems identified in the primordial-D/H database. Compilation entries were separated into their constituent absorbers, and each occurrence of an absorber in a database entry was counted; unidentified or generic entries were excluded. Absorber names are indicated for the 20 most frequently represented systems. The inset shows the same distribution on a logarithmic frequency scale, revealing the long tail of systems represented only a small number of times. A natural decline in frequency occurs after the eight most frequently represented absorbers (double-hatched histogram), which are examined individually in Figures~\ref{fig4} and \ref{fig5}.}
    \label{fig3}
\end{figure}

The repeated use of a limited observational foundation is further illustrated in Figure~\ref{fig3}, which shows the ranked frequency with which individual absorbers appear in the database after compilation lists have been split into their constituent systems. Each occurrence of an absorber in a database entry is counted, including appearances within multi-object compilations.

The resulting distribution is highly concentrated. A small number of absorbers account for a disproportionately large fraction of all object-level appearances in the literature. PKS1937--101 ($z = 3.572$) is the most frequently represented system with 258 entries, followed by Q1009+2956 (231) and HS0105+1619 (212). The next group includes Q1243+307 (149), J1558--0031 (140) and Q0913+072 (137), J1358+6522 (111) and Q2206--199 (110). These eight systems form a visually distinct high-frequency group, after which the ranked distribution declines significantly and rapidly. 

The logarithmic inset in Figure~\ref{fig3} shows that the full distribution possesses a long tail. Many absorbers occur only a handful of times, and a substantial fraction appear only once or twice. The literature is therefore characterised by a highly unequal distribution of attention: a small subset of well-studied systems is repeatedly analysed and incorporated into compilations, whereas many others make only limited contributions to the published record. The eight most frequently represented absorbers account for 1348 of the 2173 absorber mentions in the database, or 62.0 percent of the total. Extending the sample to the top 20 systems raises this to 1988 mentions, corresponding to 91.5 percent of all occurrences. Thus, although 52 distinct physical absorbers are represented overall, the published literature is dominated by a comparatively small subset of repeatedly analysed systems.

The natural break after the eighth-ranked absorber motivates our focus on these systems in the following subsections. This choice is empirical rather than imposed \textit{a priori}: the top eight constitute the group for which the combination of publication frequency, repeated analysis and long temporal baseline is sufficient to reconstruct detailed histories of measurement and reuse.

\subsection{Publication histories of the dominant absorbers}

\begin{figure}
    \centering
    \includegraphics[width=\columnwidth]{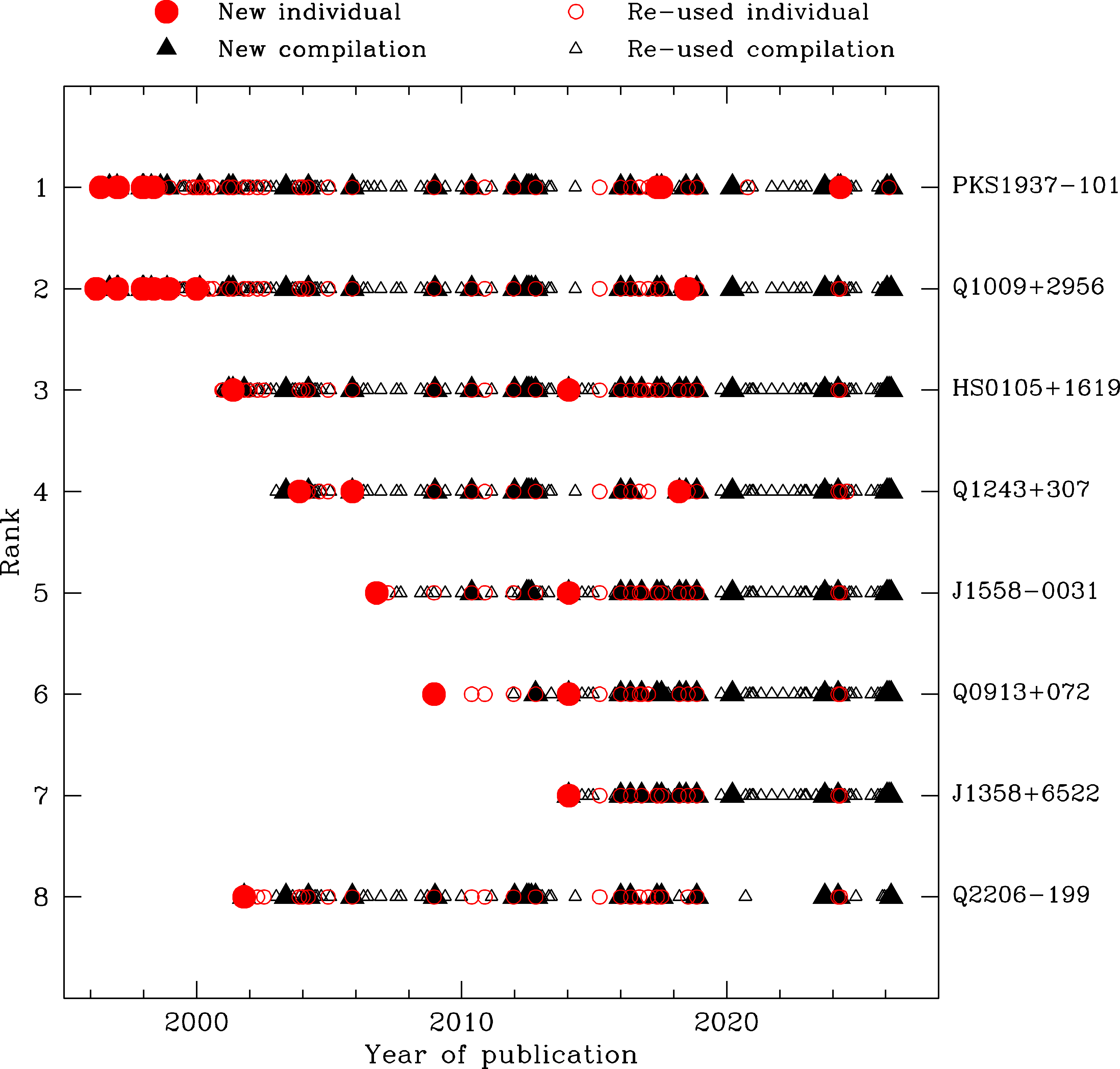}
    \caption{Publication histories of the eight most frequently represented absorption systems in the database, ranked according to their total number of appearances in Figure~\ref{fig3}. Filled red circles denote new individual observational determinations (\texttt{OBS\_INDIV\_NEW}), open red circles represent reused individual values (\texttt{OBS\_INDIV\_REV}), filled black triangles are new observational compilation values (\texttt{OBS\_COMP\_NEW}) and open black triangles display reused compilation values (\texttt{OBS\_COMP\_REV}). }
    \label{fig4}
\end{figure}

Figure~\ref{fig4} shows the publication histories of the eight most frequently represented absorbers. For each system, individual publications are classified according to whether they present a new individual determination, reuse or repeat an existing individual value, present a new compilation including that absorber or reuse an existing compilation value. Multiple database entries associated with the same absorber, source publication, category and publication month are represented by a single event. The figure therefore traces the publication-level histories of the dominant absorbers rather than the raw number of database occurrences.

Several general features are immediately apparent. First, genuinely new individual determinations are sparse compared with the total number of later appearances. In most cases, a small number of observationally new measurements is followed by a much larger number of repeated or compiled uses. The visual density of the later literature therefore reflects continued propagation of existing information at least as much as the production of new observations.

Second, the publication histories differ markedly among absorbers. Some systems enter the literature early and remain in active use for decades, whereas others appear only after the field had already entered its modern precision era. J1358+6522, for example, has a relatively short but dense publication history, whereas the use of PKS1937--101, Q1009+2956 and Q2206--199 extends over much longer baselines.

Third, new compilation values become increasingly prominent in the later literature. Once an absorber has become part of a widely adopted sample, subsequent appearances may occur predominantly through compilation-level analyses rather than through renewed individual study. In this sense, the literature develops an additional layer of dependence: not only are the same physical absorbers reused, but particular compilations become recurring means through which their values are propagated.

Figure~\ref{fig4} therefore provides a direct historical representation of the distinction between new information and repeated appearance. The total number of literature occurrences for the dominant systems is large, but the number of genuinely new observational events from which those occurrences ultimately originate is much smaller.

\subsection{Numerical evolution of the dominant absorbers}

\begin{figure}
    \centering
    \includegraphics[width=\columnwidth]{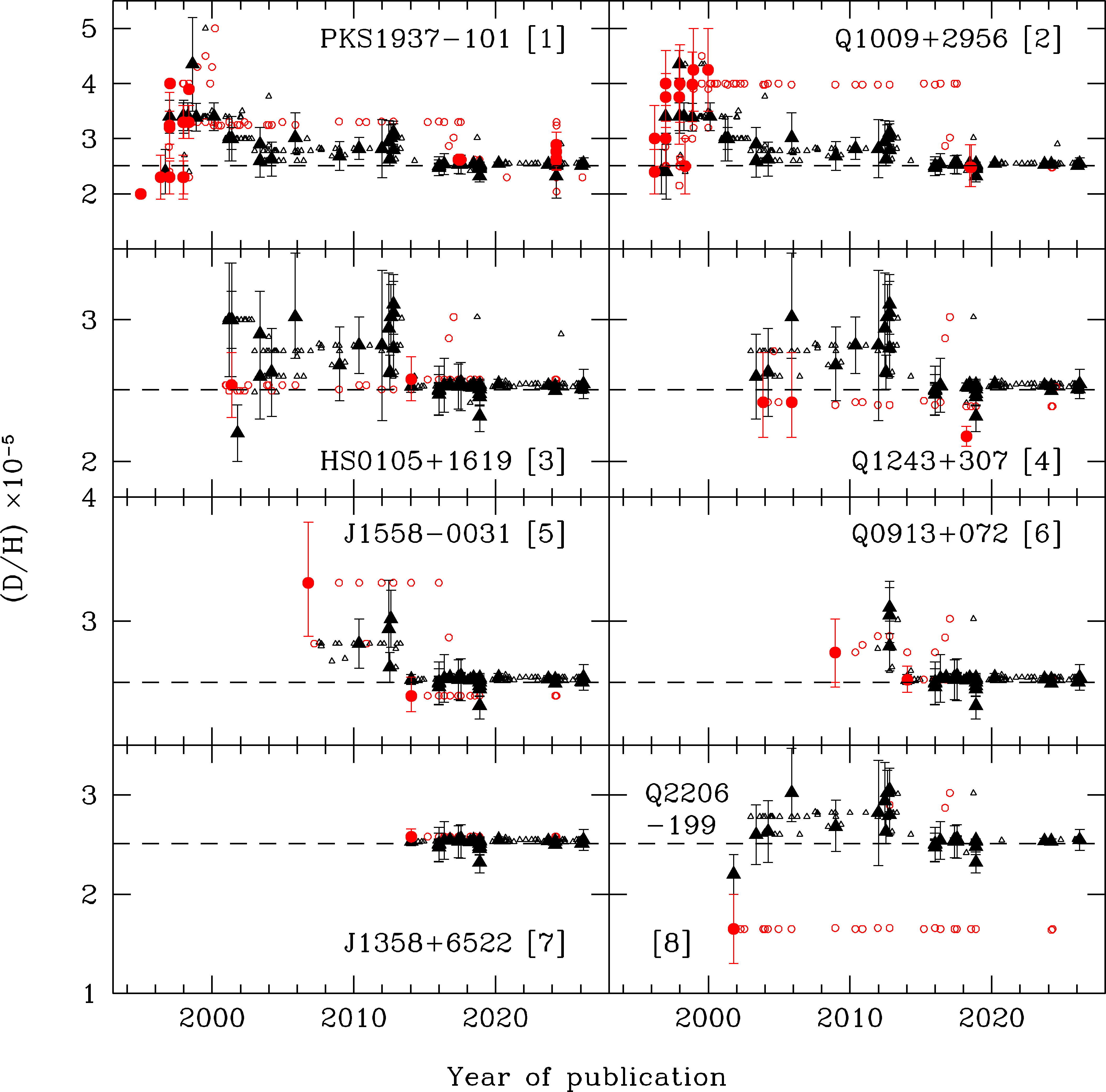}
    \caption{Evolution of the published primordial-D/H values associated with the eight most frequently represented absorption systems. Panels are ordered by frequency rank as in Figures~\ref{fig3} and \ref{fig4}, with ranks indicated in square brackets. New individual determinations and new compilation values are shown with their reported uncertainties, whereas reused individual and compilation values are shown as smaller symbols without error bars to emphasise their subsequent numerical propagation through the literature. Symbol definitions follow Figure~\ref{fig4}. The horizontal dashed line in each panel marks the PDG's 2025 recommended observational value, $10^5({\rm D/H})_{\rm p}=2.508$. }
    \label{fig5}
\end{figure}

The publication histories in Figure~\ref{fig4} establish when and how the dominant absorbers were used. Figure~\ref{fig5} shows how the numerical D/H values associated with those systems evolved. New individual measurements and new compilation values are shown with their quoted uncertainties, whereas reused values are included as smaller symbols without error bars, for clarity. The current reference abundance is shown as horizontal dashed lines. Figure~\ref{fig5} thus illustrates both the eventual convergence of the dominant systems towards the modern reference abundance and the persistence, in several cases, of previously published values offset from that locus.

The eight systems exhibit markedly different early histories but converge remarkably closely towards the modern reference value. Several absorbers show large initial offsets, whereas others were already close to the present consensus when first measured. In all cases, later new determinations and compilations occupy a much narrower range than the earliest values.

Figure~\ref{fig5} also reveals that convergence in the underlying measurements does not imply immediate disappearance of earlier values from the literature. Particularly clear examples occur for Q1009+2956, Q1243+307, J1558--0031 and Q2206--199, where sequences of reused values remain visibly offset from the modern reference level for extended periods. These (near-)horizontal sequences demonstrate that numerical values can persist in the literature even after more recent measurements have already clearly suggested a substantial revision.

A related pattern is seen in several systems where a newly published individual value is noticeably displaced from a contemporaneous compilation value, yet the compilation value subsequently becomes the one more frequently propagated through later publications. Q1243+307, J1558--0031 and Q2206--199 provide particularly clear examples of this behaviour. In such cases, compilations do not merely summarise existing information; they can become new numerical reference points within the subsequent literature.

A further feature becomes apparent when the individual and compilation categories are considered separately. For the seven dominant absorbers with publication histories extending before 2014, the compilation values show a conspicuous transition towards the modern low-D/H regime around 2013--2014. No comparably synchronous transition is evident among reused individual values, which continue to occupy a broader range. New individual determinations published subsequently are generally associated with the lower-D/H regime, although they retain substantial absorber-to-absorber variation. The coordinated change in the compilation literature is considered further in Section~\ref{discussion.sec}.

This behaviour introduces an important distinction between observational and bibliographic inheritance. A later paper may be statistically dependent on an earlier study because it uses the same absorber or source spectrum, but it may also be numerically dependent because it adopts a previously published value without independently reconstructing the underlying measurement. The resulting literature therefore possesses a genealogy in which particular numerical estimates can persist across multiple generations of publication.

The broad convergence of all eight systems towards the present reference abundance is striking. Their early values approach the modern consensus from different directions and on different timescales, yet the most recent measurements and compilations occupy a narrow common interval. This agreement is encouraging from the perspective of precision BBN, but it also motivates a more quantitative examination of the rate and structure of convergence.

The individual-measurement histories nevertheless remain heterogeneous. Some systems approach the modern abundance progressively, whereas others undergo substantial revisions following new observations or reanalysis, and reused values can preserve earlier numerical states long after newer determinations have appeared. By contrast, the compilation histories display the more coordinated 2013–2014 transition identified above. Thus, the emergence of the modern consensus involved both absorber-specific observational developments and a nearly synchronous change in the way measurements were selected, combined and propagated through the compilation literature. The publications and methodological developments associated with this transition are examined in Section~\ref{discussion.sec}.

Figures~\ref{fig1}--\ref{fig5} together reveal a consistent historical structure. The modern primordial-D/H consensus is supported by a relatively small number of physical absorption systems whose measured values have converged closely over time. Meanwhile, the published record contains extensive reuse of those systems and persistence of earlier numerical values, such that the number of reported determinations greatly exceeds the number of independent observational constraints. In the next section, we quantify the statistical consequences of this structure and examine the extent to which conventional treatments of the published literature may overstate its effective independence and precision.

\section{Statistical analysis and effective independence}
\label{statistics.sec}

The historical analysis in Section~\ref{structure.sec} demonstrates that the primordial-D/H literature is strongly hierarchical. Hundreds of published observational values ultimately derive from a much smaller number of physical absorption systems, and a small subset of those systems accounts for most appearances in the literature. Consequently, conventional statistical treatments in which individual published values are assumed to be independent risk assigning disproportionate weight to repeatedly reobserved and reanalysed absorbers. We therefore use the provenance information assembled in Section~\ref{data.sec} to examine separately the introduction of new information, the reuse of published values and the sensitivity of inferred ensemble properties to absorber-level clustering.

\subsection{New information versus repeated appearances}
\label{new_repeated.sec}

The provenance classifications introduced in Section~\ref{data.sec} allow us to quantify directly the balance between the introduction of new results and the subsequent reuse of previously published information. Of the 957 entries in the full database, 795 are observational and 162 are BBN-derived predictions. The observational subset comprises 91 new individual-system determinations (\texttt{OBS\_INDIV\_NEW}), 67 new compilation results (\texttt{OBS\_COMP\_NEW}), 456 reused individual values (\texttt{OBS\_INDIV\_REV}) and 181 reused compilation values (\texttt{OBS\_COMP\_REV}).

Repeated values therefore dominate the observational literature. Taken together, the two reused categories contain 637 entries, or 80.1 percent of all observational D/H values in our database. Only 158 entries (19.9 percent) represent newly published individual or compilation results. More restrictively, just 91 entries, or 11.4 percent of the observational database, correspond to new determinations for individual absorption systems.

This distinction is important because a new compilation does not necessarily introduce new observational information. The 67 \texttt{OBS\_COMP\_NEW} entries represent newly calculated ensemble estimates, but in most cases they combine measurements of absorbers that had already entered the literature. Hence the 91 \texttt{OBS\_INDIV\_NEW} entries provide a more direct measure of the rate at which new absorber-level observational information entered
the published record, although even these 91 entries do not correspond to 91 distinct physical systems (cf.\ Section~\ref{growth.sec}).

The imbalance is particularly pronounced among individual-system values. Of the 547 individual observational entries, only 91 (16.6 percent) are classified as new, whereas 456 (83.4 percent) represent subsequent reuse. Compilation values show the same qualitative pattern, although less extremely: 67 of 248 (27.0 percent) are new compilation results and 181 (73.0 percent) are reused compilation values. The numerical growth of the published primordial-D/H literature has therefore been driven predominantly by the repeated appearance and recombination of existing information rather than by a comparable increase in newly measured systems.

The BBN-derived subset exhibits a different pattern. Of the 162 theoretical entries in the database, 97 (59.9 percent) are classified as new predictions (\texttt{BBN\_PRED\_NEW}), whereas 65 (40.1 percent) reproduce or reuse previously published predictions (\texttt{BBN\_PRED\_REV}). Thus, in contrast to the observational literature, newly calculated values form the majority of BBN entries. This difference is not unexpected: successive BBN calculations may
incorporate updated nuclear reaction rates, cosmological parameters or computational treatments while addressing the same underlying physical quantity. Nevertheless, the substantial fraction of reused BBN values shows that numerical inheritance is not confined to the observational literature.

Considering the full database, 255 of the 957 entries (26.6 percent) are classified as new results of one kind or another, whereas 702 (73.4 percent) represent reused or repeated values. This aggregate fraction should be interpreted cautiously, because the meaning of `new' differs between individual measurements, compilations and theoretical predictions. The more informative result for assessing observational independence remains the strong dominance of reused values within the 795 observational entries, where 80.1 percent are classified as repeated individual or compilation values.

\subsection{Convergence of new and reused values}
\label{convergence.sec}

Figure~\ref{fig5} suggests that newly generated measurements and compilations tend to approach the modern primordial-D/H abundance more rapidly than values propagated from the earlier literature. To quantify this behaviour, we define for the $i$th observational value the absolute deviation from the present-day reference abundance as
\begin{equation}
\delta_i = \left|10^5({\rm D/H})_i - 2.508\right|,
\end{equation}
where $10^5({\rm D/H})_{\rm p}=2.508$ is the PDG's 2025 recommended observational value adopted throughout this paper. We emphasise that $\delta_i$ is used here solely as a retrospective diagnostic of convergence. The present-day reference value was not available to earlier studies and should not be interpreted as a value towards which historical measurements were expected to converge.

For this analysis we consider the observational entries only. Measurements reported exclusively as upper or lower limits were excluded, since treating the limiting value as a measured central abundance would give $\delta_i$ an unclear statistical meaning. This leaves 721 measured observational values: 78 new individual determinations (\texttt{OBS\_INDIV\_NEW}), 67 new compilation values (\texttt{OBS\_COMP\_NEW}), 402 reused individual values (\texttt{OBS\_INDIV\_REV}) and 174 reused compilation values (\texttt{OBS\_COMP\_REV}).

Table~\ref{tab:convergence_classes} summarises the distributions of $\delta_i$ across the full 1994--2026 interval. New individual determinations have a median absolute deviation of 0.356, compared with 0.615 for reused individual values. The corresponding medians for new and reused compilations are substantially smaller, at 0.058 and 0.122, respectively. Combining the two `new' categories gives a median $\delta =0.208$, compared with 0.312 among reused values. Thus, considered over the full history of the field, values carried forward from earlier publications are on average less closely aligned with the modern abundance than newly generated results.

\begin{table}
\centering
\caption{Absolute deviations from the present-day reference primordial-D/H abundance for measured observational entries. The interquartile range (IQR) is given by the 25th and 75th percentiles. Upper and lower limits have been excluded.}
\label{tab:convergence_classes}
\begin{tabular}{lrcc}
\hline
Category & $N$ & Median $\delta$ & IQR \\
\hline
\texttt{OBS\_INDIV\_NEW} & 78  & 0.356 & 0.112--1.242 \\
\texttt{OBS\_INDIV\_REV} & 402 & 0.615 & 0.088--1.465 \\
\texttt{OBS\_COMP\_NEW}  & 67  & 0.058 & 0.024--0.447 \\
\texttt{OBS\_COMP\_REV}  & 174 & 0.122 & 0.037--0.492 \\
\hline
All new values              & 145 & 0.208 & 0.042--0.792 \\
All reused values           & 576 & 0.312 & 0.072--0.992 \\
\hline
\end{tabular}
\end{table}

The full-period statistics mix several distinct stages in the historical development of primordial-D/H measurements. We therefore repeated the comparison within four broad periods chosen to separate the exploratory period of the 1990s, the subsequent consolidation of the observational abundance, the
rapid convergence evident during the 2010s and the recent precision era. The results are included in Table~\ref{tab:convergence_epochs}.

During 1994--2000, both newly generated and reused values remained widely dispersed relative to the modern abundance. The median absolute deviation is 0.892 for new values and 1.842 for reused values. By 2001--2010 these decrease to 0.462 and 0.492, respectively, and the difference between the two populations is small. A substantially different pattern emerges during the interval 2011--2017. Newly generated results have by then converged closely on the modern reference value, with a median $\delta =0.058$, whereas reused values retain a median deviation of 0.257, more
than four times larger. By 2018--2026 the distinction largely disappears again: the corresponding median deviations are 0.037 and 0.045.

\begin{table*}
\centering
\caption{Temporal evolution of the median absolute deviation $\delta =\left|10^5({\rm D/H})-2.508\right|$ for new and reused observational values. The final four columns separate individual measurements and compilation values. Values reported only as limits have been excluded.}
\label{tab:convergence_epochs}
\begin{tabular}{ccccccc}
\hline
Period & $N_{\rm new}$ & Median & $N_{\rm reused}$ & Median & Median & Median \\
& & $\delta_{\rm new}$ & & $\delta_{\rm reused}$ & $\delta_{\rm indiv,new}$ / $\delta_{\rm indiv,reused}$ & $\delta_{\rm comp,new}$ / $\delta_{\rm comp,reused}$ \\
\hline
1994--2000 & 44 & 0.892  & 149 & 1.842 & 1.242 / 1.992 & 0.892 / 0.892 \\
2001--2010 & 29 & 0.462  & 147 & 0.492 & 0.802 / 0.792 & 0.352 / 0.272 \\
2011--2017 & 38 & 0.058  & 146 & 0.257 & 0.110 / 0.297 & 0.041 / 0.039 \\
2018--2026 & 34 & 0.037  & 134 & 0.045 & 0.118 / 0.112 & 0.028 / 0.037 \\
\hline
\end{tabular}
\end{table*}

The origin of the pronounced 2011--2017 difference becomes clearer when individual and compilation values are considered separately. For individual absorption systems, new measurements published during this period have a median deviation of 0.110, compared with 0.297 for reused individual values. By contrast, the compilation literature had already converged much more tightly: the median deviations of new and reused compilation values are 0.041 and 0.039, respectively. Thus, the persistence of displaced values during this period is primarily associated with the continued circulation of earlier individual-system measurements rather than with compilation values as a whole.

This result provides quantitative support for the behaviour seen in Figure~\ref{fig5}. In several of the dominant absorbers, older numerical determinations continue to appear in the literature after newer measurements or compilations have shifted substantially closer to the modern abundance. Conversely, once tightly converged compilation values became established, those values themselves could subsequently be reused, explaining why the distinction between new and reused compilations becomes small during the later periods. Numerical inheritance therefore need not preserve an outdated value indefinitely: the values being inherited can themselves change as the literature develops.

The improvement in agreement is expected to reflect genuine advances in spectral quality, absorption-line modelling, sample selection and the control of systematic uncertainties. The purpose of the comparison is to distinguish the temporal behaviour of newly generated results from that of values propagated through the existing literature. From this perspective, the 2011--2017 interval is particularly revealing: the new literature had largely moved towards the modern consensus while a substantial population of older individual-system values remained in active circulation.

Finally, the 721 entries analysed here cannot be regarded as 721 statistically independent measurements. Reused values are dependent by construction, whereas different entries classified as new may still refer to the same physical absorber or share observational and methodological information. We therefore treat the distributions in Tables~\ref{tab:convergence_classes} and \ref{tab:convergence_epochs} as descriptive measures of the published record rather than as independent samples to which conventional row-by-row significance tests can be applied. The effects of absorber-level clustering and the resulting reduction in effective statistical independence are examined in the next subsection.

\subsection{Absorber-level clustering and effective representation}
\label{clustering.sec}

The analysis above treats the individual entries in the database as features of the published literature, but it does not imply that these entries constitute statistically independent measurements. The provenance reconstruction in Section~\ref{data.sec} and the frequency distribution in Figure~\ref{fig3} show that the observational literature is strongly clustered at the level of individual absorption systems. Repeated analyses, revisions and compilations cause some absorbers to appear hundreds of times, whereas many others occur only once or a few times.

Across the 52 distinct physical absorbers identified in the database, there are 2173 identifiable absorber mentions. Their distribution is highly unequal. The eight most frequently represented systems account for 1348 mentions, or 62.0 percent of the total, and the top 20 account for 1988 mentions, or 91.5 percent. Under equal representation, these fractions would be only 15.4 and 38.5 percent, respectively. Hence, simple counting of database entries assigns substantially greater implicit weight to a small subset of repeatedly analysed systems.

To quantify this concentration independently of the D/H values themselves, we use the inverse of the Herfindahl--Hirschman concentration index \citep{Herfindahl1950, Hirschman1964} to define an effective number of represented absorbers,
\begin{equation}
N_{\rm eff} = \left(\sum_{j=1}^{N_{\rm abs}} p_j^2\right)^{-1},
\end{equation}
where $p_j=n_j/\sum_k n_k$ is the fractional representation of absorber $j$ among all identifiable absorber mentions. The Herfindahl--Hirschman index itself is given by $H=\sum_j p_j^2$, such that $N_{\rm eff}=H^{-1}$. When all absorbers are represented equally, $N_{\rm eff}$ is identical to the actual number of absorbers; increasing concentration among a smaller number of systems reduces $N_{\rm eff}$. For the 52 absorbers in our database, we obtain $N_{\rm eff}=16.28$. Thus, in terms of representation frequency, the observed literature has the same degree of concentration as a sample containing only approximately 16 equally represented absorbers.

\begin{table}
\centering
\caption{Concentration of identifiable absorber mentions in the primordial-D/H literature. The expected fractions assume equal representation of all 52 absorbers.}
\label{tab:absorber_concentration}
\begin{tabular}{lrcr}
\hline
Subset & Absorbers & Mentions & Fraction (\%) \\
\hline
Top 8  & 8  & 1348 & 62.0 \\
Top 20 & 20 & 1988 & 91.5 \\
All    & 52 & 2173 & 100.0 \\
\hline
\multicolumn{3}{l}{Effective number, $N_{\rm eff}$} & 16.28 \\
\hline
\end{tabular}
\end{table}

Because the statistic is cumulative, older absorbers have naturally had more opportunities to recur in later measurements and compilations. We therefore tested whether the inferred concentration is primarily a consequence of publication age by recalculating the inverse Herfindahl--Hirschman index over restricted recent publication windows. Restricting the analysis to absorber appearances from 2012 onwards yields an effective representation of $\simeq 16.8$ equally represented absorbers, essentially unchanged from the full-sample value. For the 2018--2026 literature the corresponding value is about 14.5, among only 22 absorbers represented during that interval. Thus, the strong concentration of the literature is not simply a consequence of older systems having had more time to accumulate appearances; it remains evident, and if anything becomes stronger, in the recent literature.

These quantities should nevertheless be interpreted purely as measures of bibliographic concentration. The fractions $p_j$ describe the representation
of physical absorbers in the published D/H literature, rather than conventional citation counts or statistical weights, and the resulting $N_{\rm eff}$ is not a formal estimate of the number of statistically independent measurements. We emphasise that $N_{\rm eff}$ does not incorporate the quoted measurement uncertainties, the degree of covariance among analyses of the same spectra or methodological correlations between or among different absorbers. Nevertheless, it illustrates the scale of the weighting problem: although 52 physical systems can be identified, the numerical literature is dominated by a much smaller effective subset.

This distinction is particularly important for inverse-variance weighting. If every published value is treated as an independent measurement, an absorber that has been measured, reanalysed or incorporated into compilations many times contributes repeatedly to the likelihood or weighted mean. Its influence then depends not only on the precision of the underlying observations but also on the frequency with which that system has subsequently appeared in the literature. In the limiting case in which several publications merely repeat the same numerical result and uncertainty, treating these entries as independent would reduce the formal uncertainty without adding any new observational information.

An absorber-level treatment addresses this problem by regarding measurements associated with the same physical system as belonging to a common statistical cluster. Repeated analyses can then improve our understanding of that absorber, but they should not increase the number of independent sightlines through which primordial D/H is sampled. This distinction is analogous to the dataset-level correlations identified in our analysis of primordial helium abundances \citep{2026PASA...43...82H}, but the present database permits the dependence structure to be traced more directly because the underlying absorption systems can usually be identified explicitly.

\subsection{Sensitivity of the inferred abundance to repeated absorbers}
\label{sensitivity.sec}

The strong concentration of the literature quantified in Section~\ref{clustering.sec} raises a practical question: to what extent would a conventional combination of published D/H measurements depend on the absorbers that appear most frequently? We address this using both leave-one-absorber-out tests and several alternative absorber-level summaries. The purpose of these calculations is not to derive a new preferred primordial-deuterium abundance but to quantify how strongly an apparently precise literature-level estimate depends on the statistical representation of repeated measurements of the same physical systems.

For the primary analysis we define a modern observational sample comprising measured values classified as \texttt{OBS\_INDIV\_NEW} and published from 2012 onwards. This boundary corresponds approximately to the onset of the more tightly converged regime seen in Figures~\ref{fig1} and \ref{fig5}, while avoiding the larger scatter still evident immediately beforehand. The resulting sample contains 29 measurements associated with 15 distinct absorption systems. Their representation remains highly unequal. The two absorbers towards PKS1937--101 together contribute 13 of the 29 measurements, 11 of which are associated with the $z=3.572$ absorber. We verified that shifting the lower boundary to 2011 or 2013 has little effect on the principal conclusions below.

As a deliberately naive baseline, we first treat all 29 published values as statistically independent and calculate the standard inverse-variance weighted mean,
\begin{equation}
\bar{x}_{\rm w} = \frac{\sum_i x_i/\sigma_i^2}{\sum_i 1/\sigma_i^2},
\end{equation}
with formal uncertainty
\begin{equation}
\sigma_{\bar{x}_{\rm w}} = \left(\sum_i \sigma_i^{-2}\right)^{-1/2}.
\end{equation}
Using the measurement uncertainties quoted in the literature gives
\begin{equation}
10^5({\rm D/H}) = 2.5577 \pm 0.0150.
\end{equation}
This lies 0.0497 above the PDG's 2025 reference value, $10^5({\rm D/H})_{\rm p}=2.508$. The calculation is intentionally diagnostic: by treating all 29 entries as independent, it ignores precisely the absorber-level clustering established in the preceding sections.

To quantify the sensitivity of this result to individual physical systems, we repeated the calculation after removing in turn all entries associated with each of the 15 absorbers. The resulting weighted means cover $2.4718 \leq 10^5({\rm D/H}) \leq 2.5761$. The largest downward shift occurs when the $z=3.572$ absorber towards PKS1937--101 is excluded, reducing the weighted mean from 2.5577 to 2.4718. Conversely, exclusion of Q1243+307 produces the largest upward shift, to 2.5761.

The sensitivity to PKS1937--101 is particularly instructive. Removal of this single physical absorber changes the inferred abundance by 0.0860, almost six times the formal $1\sigma$ uncertainty of the full 29-entry weighted mean. The apparent precision of the naive literature-level estimate therefore substantially understates its sensitivity to the absorber-level structure of the data. The effect is amplified by the fact that the main PKS1937--101 absorber contributes 11 separate entries and consequently receives repeated statistical weight.

A complementary test is obtained by excluding collectively the eight absorbers that dominate the frequency distribution in Figure~\ref{fig3}. The remaining modern sample contains 11 measurements associated with eight absorbers and yields
\begin{equation}
10^5({\rm D/H}) = 2.4960 \pm 0.0265,
\end{equation}
again under the assumption that the retained measurements are independent. This value differs by only $-0.0120$ from the adopted 2025 PDG abundance. These principal absorber-selection tests are summarised in Table~\ref{tab:sensitivity}.

This experiment addresses a different robustness question from the leave-one-absorber-out analysis. The leave-one-out calculation asks how strongly the naive weighted mean responds to the removal of any single physical system and hence diagnoses the leverage of individual absorbers. By contrast, removing all eight dominant systems asks whether the modern consensus near $10^5({\rm D/H})\simeq2.5$ remains visible in the less frequently represented part of the observational sample. The answer is affirmative: although the dominant absorbers can exert substantial leverage on a publication-level weighted mean, the consensus itself is not confined to that high-frequency subset.

\begin{table*}
\centering
\caption{Sensitivity of the inverse-variance weighted mean of the modern \texttt{OBS\_INDIV\_NEW} sample, defined as measurements published from 2012 onwards, to absorber-level selection. The quoted uncertainties are formal errors obtained by treating the retained entries as independent and are therefore used only as diagnostics. The final column gives the offset from the adopted 2025 PDG value, $10^5({\rm D/H})_{\rm p}=2.508$.}
\label{tab:sensitivity}
\begin{tabular}{lccc}
\hline
Sample & $10^5({\rm D/H})$ & Formal uncertainty & $\Delta_{\rm PDG}$ \\
\hline
All 29 entries & 2.5577 & 0.0150 & +0.0497 \\
Exclude PKS1937--101 ($z = 3.572$) & 2.4718 & 0.0224 & $-0.0362$ \\
Exclude Q1243+307 & 2.5761 & 0.0154 & +0.0681 \\
Exclude all top-eight absorbers & 2.4960 & 0.0265 & $-0.0120$ \\
\hline
\end{tabular}
\end{table*}

The leave-one-absorber-out experiment diagnoses the sensitivity of the value-level weighted mean, but it does not itself remove the unequal representation of individual physical systems. We therefore also considered several simple absorber-level summaries of the same modern sample. These calculations are again intended as sensitivity tests rather than alternative determinations of the primordial deuterium abundance.

As a first, deliberately robust approach, all \texttt{OBS\_INDIV\_NEW} measurements associated with a given absorber were collapsed to the median D/H value for that system. The resulting 15 absorber-level values have a median
\begin{equation}
10^5({\rm D/H}) = 2.5000,
\end{equation}
with a median absolute deviation of 0.100. This central value lies only 0.008 below the adopted 2025 PDG abundance. A non-parametric bootstrap in which the 15 absorbers, rather than the individual database entries, are resampled gives an approximate 95 percent interval of 2.389--2.580 for the median. The breadth of this interval reflects both the limited number of distinct systems and the remaining absorber-to-absorber dispersion; it should not be compared directly with the formal inverse-variance uncertainty, since the quantities have different statistical interpretations.

For comparison, the equal-weight arithmetic mean of the 15 absorber-level medians is
\begin{equation}
10^5({\rm D/H}) = 2.465 \pm 0.051,
\end{equation}
where the quoted uncertainty is the standard error of the mean across absorbers. The lower central value is influenced by several relatively low D/H determinations with comparatively large observational uncertainties, illustrating the limitations of assigning identical weight to measurements of very different precision. We therefore do not regard either the equal-weight mean or the median as a preferred estimator. Their role is to show how the inferred ensemble abundance changes when the physical absorber, rather than the individual published measurement, is taken as the fundamental statistical unit.

As a second absorber-level sensitivity test, we selected for each of the 15 absorbers the \texttt{OBS\_INDIV\_NEW} determination with the smallest quoted uncertainty. For measurements with asymmetric uncertainties, we adopted $\sigma=(\sigma_{+}+\sigma_{-})/2$ for the purpose of this selection and subsequent inverse-variance weighting. This restricts each physical absorber to a single contribution to the calculation while retaining information about the quoted measurement precision. The resulting weighted mean is
\begin{equation}
10^5({\rm D/H}) = 2.5316 \pm 0.0196.
\end{equation}
This lies 0.0236 above the adopted PDG abundance.

Selection of the smallest quoted uncertainty per absorber is itself an analyst-defined rule, however, and may favour measurements with particularly small reported errors. This estimate should therefore also be interpreted as a diagnostic rather than as a new recommended primordial-D/H abundance. The contrast among the various treatments is summarised in Table~\ref{tab:absorber_estimators}.

\begin{table*}
\centering
\caption{Sensitivity of summary primordial-D/H estimates to the choice of statistical unit for the modern \texttt{OBS\_INDIV\_NEW} sample. The final column gives the offset from the adopted 2025 PDG value, $10^5({\rm D/H})_{\rm p}=2.508$. These estimates are diagnostic comparisons and are not intended as new recommended determinations of the primordial D/H abundance. MAD: Mean absolute deviation.}
\label{tab:absorber_estimators}
\begin{tabular}{lccc}
\hline
Treatment & $10^5({\rm D/H})$ & Uncertainty/dispersion & $\Delta_{\rm PDG}$ \\
\hline
29 value-level measurements & 2.5577 & 0.0150 & +0.0497 \\
15 absorbers, smallest uncertainty & 2.5316 & 0.0196 & +0.0236 \\
15 absorber medians, equal-weight mean & 2.465 & 0.051 & $-0.043$ \\
15 absorber medians, median & 2.5000 & MAD $=0.100$ & $-0.0080$ \\
\hline
\end{tabular}
\end{table*}

\subsection{Robustness of the modern consensus}
\label{robust.sec}

The analyses in Sections~\ref{new_repeated.sec}--\ref{sensitivity.sec} show that two questions must be distinguished when assessing the modern primordial-D/H literature: the robustness of the central abundance and the statistical independence underlying its quoted precision.

The provenance analysis demonstrates that the published record is strongly dependent. Reused observational values constitute 80.1 percent of the observational database, whereas only 11.4 percent of all observational entries correspond to new determinations of individual absorption systems. The absorber-frequency distribution is similarly concentrated: although 52 distinct physical systems are represented, the inverse Herfindahl--Hirschman measure corresponds to only $N_{\rm eff}=16.28$ equally represented absorbers. Thus, the size of the published literature greatly exceeds the effective breadth of its underlying observational foundation.

The temporal analysis nevertheless shows that numerical reuse does not simply preserve obsolete values indefinitely. During the main convergence phase, particularly in the period 2011--2017, newly generated values lie substantially closer to the present-day reference abundance than reused values. By 2018--2026 this difference has largely disappeared, indicating that once the literature converged, the values subsequently propagated through reuse were themselves close to the modern consensus.

The sensitivity tests provide the most direct assessment of whether this dependence compromises the modern central abundance. When all 29 \texttt{OBS\_INDIV\_NEW} measurements published from 2012 onwards are treated as independent, the inverse-variance weighted mean lies above the adopted 2025 PDG abundance. However, alternative treatments that reduce or remove repeated absorber-level representation continue to recover values close to $10^5({\rm D/H})\simeq2.5$. In particular, the median of the 15 absorber-level medians is $10^5({\rm D/H})=2.5000$, only 0.0080 below the adopted 2025 PDG value of 2.508. Likewise, removal of all eight absorbers that dominate the frequency distribution yields $10^5({\rm D/H})=2.4960\pm0.0265$ for the remaining modern measurements.

These results indicate that the modern consensus is not an artefact of one or two repeatedly analysed absorbers, nor is it confined to the small group of systems that dominate the published literature. The convergence towards $10^5({\rm D/H})\simeq2.5$ remains visible when those dominant systems are summarised at absorber level, represented only once or removed altogether.

The formal precision of the literature-level constraint is less straightforward. Repeated measurements of the same physical absorber can improve the determination for that system, but they do not create additional independent sightlines. Treating such entries as independent therefore allows publication frequency to influence the statistical weight assigned to a physical system and can produce formal uncertainties that are smaller than the absorber-level
dependence structure warrants. Compilation values introduce an additional layer of numerical dependence because they may repeatedly propagate combinations of the same underlying measurements.

Consequently, the robustness of the central primordial-D/H abundance and the robustness of its formal statistical uncertainty should be regarded as separate questions. Our analysis provides strong support for the former: a value close to the adopted 2025 PDG abundance is recovered under several substantially different treatments of the observational genealogy. The formal statistical uncertainty is more sensitive to how repeated measurements and shared absorbers are
represented statistically. The modern primordial-D/H constraint therefore appears robust in its central value but less independent, and hence less straightforwardly precise, than a simple count of published measurements would imply.

The statistical analysis establishes this empirical structure but does not explain why particular absorber values persisted, shifted abruptly or became dominant in later compilations. Those questions require examination of the observational and methodological histories of the individual systems. We turn to these processes in Section~\ref{discussion.sec}.

\section{Discussion}
\label{discussion.sec}

\subsection{From measurement to consensus}

The historical evolution reconstructed in Section~\ref{structure.sec} shows that the modern primordial-D/H abundance did not emerge simply through the
accumulation of an increasing number of measurements. The early literature was characterised by large dispersion, substantial uncertainties and, in some cases, mutually inconsistent determinations. Over time, the range of reported values narrowed substantially and the field moved towards the tightly clustered modern regime near $10^5({\rm D/H})\simeq2.5$.

This transition reflects a change, not just in the quantity of available data but also in the nature of the observational problem itself. Early work established the feasibility of measuring deuterium in high-redshift absorption systems while simultaneously revealing the principal sources of ambiguity, including contamination by unrelated H{\sc i} absorption and the difficulty of modelling complex velocity structures. Subsequent progress increasingly relied on high-resolution, high-signal-to-noise spectroscopy combined with more stringent selection of low-metallicity systems with simple, well-resolved absorption structure. The result was a gradual shift from an exploratory regime, in which a wide range of primordial-D/H values remained observationally plausible, towards one in which a small subset of particularly favourable absorbers could be measured with much greater precision.

The emergence of consensus should therefore not be understood as a pure averaging process. It reflects a progressive refinement of what constitutes a reliable primordial-D/H measurement. Improvements in instrumentation, absorption-line modelling and the treatment of systematic uncertainties reduced the dispersion of individual determinations, whereas the growing experience of the community made it possible to identify which absorption systems were most suitable for precision work. In this sense, the history of primordial D/H is one of increasing selectivity as much as increasing precision.

This distinction is important because the observational foundation did not expand at the same rate as the literature built around it. As shown in Section~\ref{growth.sec}, the number of publications and reported values grew much more rapidly than the number of newly introduced absorption systems. The modern consensus consequently emerged from repeated scrutiny of a relatively small observational foundation rather than from the continual addition of large numbers of independent sightlines. The convergence seen in Figures~\ref{fig1}--\ref{fig5} is thus best interpreted as the combined result of improved measurement practice, increasingly selective use of favourable systems and the reanalysis and synthesis of an established absorber population. By the early 2010s, this process increasingly involved the construction of deliberately restricted samples selected for their suitability for precision D/H work and analysed within increasingly homogeneous frameworks.

The close agreement between modern observational determinations and the adopted 2025 PDG abundance is therefore scientifically meaningful. At the same time, the route by which that agreement was reached matters for its statistical interpretation. A tightly clustered literature can arise because measurements themselves have improved, the same successful systems have been studied repeatedly or particular values and compilations have become widely propagated. The analyses in this paper indicate that all three processes have contributed to the present-day appearance of consensus.

\subsection{The persistence of published values}

Convergence of the underlying measurements does not imply that older numerical values disappear immediately from the literature. Figure~\ref{fig5} shows several clear examples in which previously published D/H values continue to recur after newer determinations have moved closer to the modern abundance. The statistical comparison in Section~\ref{convergence.sec} quantifies the same behaviour: during the principal convergence phase,
newly generated values were typically closer to the present-day reference abundance than values propagated from earlier publications.

This persistence is a natural consequence of cumulative scientific practice. Published numerical values can continue to be used for many reasons after the original observational context has changed. Reviews and theoretical studies may quote established measurements as representative observational constraints; later compilations may retain previously adopted values for consistency; and papers concerned with questions other than the detailed measurement of D/H may rely on values already established in the literature rather than reconstruct the full observational provenance of each absorber. Once a particular number has entered common use, it can therefore remain visible across multiple generations of publications.

The resulting persistence should not be interpreted by itself as evidence of publication bias, sociological conformity or inappropriate selection of results. Reuse is an essential part of cumulative science, and a repeated value may remain entirely appropriate for the purpose for which it is cited. The important point for the present analysis is statistical rather than behavioural: a numerical value can acquire a publication history that is much longer than the sequence of independent observations from which it originated.

Our database allows this distinction to be made explicitly. An absorber may be measured only a small number of times while its D/H value appears repeatedly in later individual discussions or compilations. In such cases, the literature contains both \textit{observational inheritance}, arising from repeated use of the same physical system or spectrum, and \textit{numerical inheritance}, arising from the direct propagation of a previously published value. The two forms of dependence overlap but are not identical. A later analysis may return to the same absorber and derive a genuinely new numerical result, whereas a different paper may reproduce an earlier value without introducing new observational information.

The temporal behaviour identified in Section~\ref{convergence.sec} also shows that numerical inheritance is not static. During the period 2011--2017, reused
individual-system values remained substantially more displaced from the modern reference abundance than newly generated values, whereas the compilation literature had already converged much more closely. The object-level histories in Figure~\ref{fig5} show that this difference is not merely statistical: the compilation values undergo a much more coordinated transition than the reused individual determinations. By 2018--2026, the distinction between new and reused values had largely diminished. This indicates that the numbers being inherited change as the field evolves: once a revised abundance becomes established, it can itself become the value propagated through later publications.

Persistence is therefore better understood as a property of the genealogy of the literature than as simple resistance to scientific change. Older values can remain active for extended periods, but they may eventually be replaced by newer reference points that themselves enter repeated circulation. The specific transitions by which this occurred for individual absorbers, and the publications associated with those changes, are examined in the following subsection.

\subsection{A synchronised transition in the compilation literature}

The object-level histories in Figure~\ref{fig5} reveal a feature that is less obvious in the aggregate evolution shown in Figure~\ref{fig1}. For the seven
dominant absorbers whose publication histories extend to epochs before 2014, the compilation values undergo a conspicuous transition towards the modern low-D/H regime over a remarkably narrow interval around 2013--2014. J1358+6522 cannot provide an independent test of this transition because its D/H history begins only in 2014. The common timing among the other seven systems is nevertheless striking.

Importantly, the transition is specific primarily to the compilation literature. Reused individual-system values (\texttt{OBS\_INDIV\_REV}) remain considerably more heterogeneous, with earlier determinations continuing to circulate well after 2014. Conversely, genuinely new individual measurements published after the transition generally occupy the lower-D/H regime, although they do not simply reproduce a single canonical abundance. In systems such as Q1243+307 and J1558--0031, later new determinations can lie appreciably below the present-day reference value. The transition therefore represents convergence towards a new observational regime rather than numerical locking onto one prescribed value.

The origins of this change can be traced to developments during the preceding two years. Compilation estimates published in 2012 remained sensitive to the composition of the absorber sample. \citet{2012MNRAS.426.1427O}, for example, explicitly explored the effect of excluding two potentially problematic systems and obtained significantly different combined abundances depending on the adopted sample. Other contemporary estimates likewise used different subsets of the available absorbers and produced values spanning a relatively broad range. A lower compilation branch was already present, including the value near $10^5({\rm D/H})\simeq2.6$ associated with \citet{2012MNRAS.425.2477P}, while higher values near 2.8--3.0 continued to appear through 2013.

\citet{2012MNRAS.425.2477P} were already explicit about the underlying problem. Their new determination for J1419+0829, $10^5({\rm D/H})=2.535\pm0.05$, was obtained from a particularly favourable metal-poor damped Lyman-$\alpha$ absorber using an analysis designed to account carefully for both random and systematic uncertainties. More generally, they argued that the excess scatter among published D/H measurements could reflect unaccounted systematic effects and that further progress required a homogeneous set of suitable data analysed in a self-consistent manner.

That programme was implemented explicitly by \citet{2014ApJ...781...31C}. Rather than combining the available literature measurements indiscriminately, they defined a `Precision Sample' using strict requirements on H{\sc i} column-density determination, the availability of resolved and apparently uncontaminated D{\sc i} transitions, spectral resolution and signal-to-noise ratio, and the information required to constrain the velocity structure. Four qualifying literature systems were reanalysed using the same framework as the newly measured absorber J1358+6522. The analysis was performed blind, used purpose-built profile-fitting software and treated the dominant systematic uncertainties consistently across the sample.

The effects were absorber dependent. For J1419+0829, reanalysis changed the previously published result by only 0.005 dex. By contrast, J1558--0031 shifted from a literature value of $\log({\rm D/H})=-4.48\pm0.06$ to $-4.619\pm0.026$; smaller revisions occurred for HS0105+1619 and Q0913+072. Despite these differing individual histories, the five consistently analysed systems showed much closer mutual agreement and yielded the weighted mean
\begin{equation}
10^5({\rm D/H})_{\rm p}=2.53\pm0.04.
\end{equation}
\citet{2014ApJ...781...31C} emphasised that this central value was not markedly different from some earlier estimates; the principal advance was the substantially improved precision obtained from a deliberately selected and homogeneously analysed sample.

The synchronised transition seen in Figure~\ref{fig5} should therefore not be interpreted as seven independent absorbers undergoing simultaneous physical or observational revisions. For three of the seven systems (HS0105+1619, Q0913+072 and J1558--0031) the \citet{2014ApJ...781...31C} Precision Sample provides a direct route to the new compilation value. For PKS1937--101, Q1009+2956, Q1243+307 and Q2206--199, which were not included in that sample, the lower compilation branch had already appeared through earlier work, particularly \citet{2012MNRAS.425.2477P}, and subsequently displaced the older higher-valued compilation lineage. The common visual transition around 2013--2014 thus represents a broader change in compilation practice: increasingly stringent sample selection and homogeneous analysis established a lower and substantially tighter reference regime, while older individual determinations continued to circulate alongside it.

\subsection{Compilations as nodes of numerical inheritance}
\label{inheritance.sec}

The 2013--2014 transition illustrates that compilations can play a more consequential role than simply summarising existing measurements. Once a combined abundance becomes widely accepted, subsequent papers may adopt that value directly, often without reconstructing the individual measurements from which it was derived. A compilation can therefore become a new node in the numerical genealogy of the literature.

The years immediately preceding the modern precision regime provide a clear example. In 2012, several newly calculated compilation values coexisted despite being derived from substantially overlapping absorber populations. Differences in sample membership and in the treatment or exclusion of individual systems produced combined abundances ranging from the lower values already approaching the modern consensus to values near $10^5({\rm D/H})\simeq3$. The alternative combinations considered by \citet{2012MNRAS.426.1427O} are particularly instructive because they demonstrate directly how the ensemble abundance could change when individual absorbers regarded as problematic were included or omitted.

Lower values nevertheless entered circulation before they became dominant. Some publications adopted or reviewed the lower compilation results without introducing a new analysis of the underlying systems. At the same time, higher values traced to earlier compilations and PDG summaries remained in active use through 2013. The literature consequently contained several parallel numerical lineages rather than a single continuously evolving estimate.

The Precision Sample of \citet{2014ApJ...781...31C} changed this structure. Its importance lay not simply in the numerical result, $10^5({\rm D/H})_{\rm p}=2.53\pm0.04$, but in the combination of explicit quality criteria, common treatment of the selected systems and substantially reduced dispersion. Once established, this value was repeatedly propagated through subsequent compilation-level publications. In our database it therefore appears first as a new compilation result and subsequently as a recurring reused compilation value. The statistical identity of the later entries is different from their bibliographic identity: several publications may quote the value, but the underlying numerical information ultimately traces back to the same restricted set of systems and common analysis.

This behaviour explains why the compilation and individual-value histories need not evolve synchronously. A new compilation can rapidly replace an older field-level reference value even while previous individual-system measurements remain in circulation. Indeed, the persistence of heterogeneous \texttt{OBS\_INDIV\_REV} values after the compilation transition shows that the literature was not simply purged of discrepant measurements. Rather, a new
higher-level numerical reference point emerged and increasingly became the quantity propagated when authors required a representative primordial-D/H abundance.

Compilations therefore perform two related functions. Scientifically, they combine and evaluate the available observational evidence. Bibliographically, they can mediate the subsequent transmission of that evidence by becoming preferred numerical reference points. These functions are entirely legitimate components of cumulative science, but they generate dependencies that are easily overlooked if each appearance of a compilation value is treated as a statistically independent determination. The distinction between a measurement being repeatedly confirmed and a numerical summary being repeatedly quoted is central to interpreting the apparent density of the modern primordial-D/H literature.

\subsection{How independent is the modern primordial-D/H constraint?}

The provenance reconstruction developed here distinguishes several different meanings of independence that are often conflated in literature-level analyses. Separate publications are not necessarily independent measurements; separate analyses of the same absorber are not independent physical sightlines; and a compilation containing several absorbers may itself share most or all of its observational content with earlier compilations. Conversely, statistical dependence of this kind does not imply that the resulting central abundance is unreliable.

The modern D/H literature illustrates all of these distinctions. Although 52 distinct physical absorbers can be identified in the database, their representation is extremely unequal. The top eight systems alone account for 62.0 percent of identifiable absorber mentions, and the inverse Herfindahl--Hirschman measure corresponds to only approximately 16 equally represented absorbers. At the same time, repeatedly analysed absorbers may benefit from new spectra, improved profile modelling and better treatment of systematics. Repeated analyses can therefore add genuine information without adding a new independent line of sight.

The Precision Sample introduced by \citet{2014ApJ...781...31C} provides a useful illustration of the distinction. Its five damped Lyman-$\alpha$ systems were physically independent sites, and the authors deliberately analysed them in a common framework. In this sense the five systems provide five
independent astrophysical sightlines, even though their analyses are methodologically correlated through common selection criteria, software and modelling assumptions. Later publications that reuse the resulting weighted mean do not add further independent sightlines, even though they increase the number of literature appearances of that abundance.

Our sensitivity tests show that the modern consensus survives this dependence structure. Treating all new individual measurements published from 2012 onwards as independent gives a result whose central value is strongly sensitive to the representation of particular absorbers, whereas its formal uncertainty does not capture that absorber-level sensitivity. Reducing each absorber to a single contribution, using absorber-level medians or excluding the eight most frequently represented systems changes the numerical estimate and generally increases its uncertainty but does not eliminate the concentration of modern measurements near $10^5({\rm D/H})\simeq2.5$.

The relevant conclusion is therefore not that the modern primordial-D/H abundance rests on an illusory consensus. Rather, its central value is supported more robustly than the raw number of published values might imply, while its effective statistical independence is smaller than that number suggests. Robustness of the central abundance, independence of physical sightlines and independence of published numerical estimates are distinct properties and should not be inferred from one another.

This distinction also clarifies why provenance matters even in a mature precision literature. A conventional weighted mean may be numerically stable while still assigning excessive statistical leverage to repeatedly analysed systems. Conversely, an absorber-level analysis may retain a similar central value while yielding a more realistic impression of the breadth of the observational evidence. Provenance-aware treatment is therefore principally a correction to the interpretation of precision and independence, not an argument for abandoning the modern primordial-D/H abundance.

\subsection{Implications for precision cosmology and future measurements}

Primordial deuterium is now sufficiently precise to contribute meaningfully to tests of BBN and the cosmological baryon density. This makes the distinction between measurement precision and observational independence increasingly important. Once individual D/H determinations reach percent-level precision, further reduction of the formal uncertainty cannot be assessed solely by counting the number of values available for combination.

The history reconstructed here suggests that much of the major improvement during the past decade arose from a transition towards carefully selected absorption systems and increasingly homogeneous analysis rather than from a rapid increase in the number of suitable sightlines. This development was scientifically productive. \citet{2012MNRAS.425.2477P} explicitly identified unaccounted systematics and heterogeneous analysis as possible sources of the
earlier excess scatter, and \citet{2014ApJ...781...31C} demonstrated that a strictly selected and consistently analysed sample could yield mutually compatible measurements with substantially improved precision.

The next gains in cosmological information need not follow the same route indefinitely. Reanalysing established absorbers remains useful, particularly where improved spectra or modelling can reduce systematic uncertainties, but repeated study of the same systems cannot increase the number of independent lines of sight. The discovery and high-quality observation of additional suitable low-metallicity absorbers therefore remain especially valuable. New systems test whether the abundance and dispersion inferred from the best-studied absorbers generalise to the broader population and reduce the dependence of ensemble constraints on a small number of exceptionally favourable objects.

Future compilations would also benefit from making their provenance structure explicit. At minimum, combined primordial-D/H estimates should identify the physical absorbers contributing to the result, distinguish new measurements from reanalyses of existing systems and indicate when numerical values are inherited directly from previous compilations. Where multiple measurements of the same absorber are included, their statistical relationship should be considered rather than assuming independence by default. Such information would make it possible to distinguish increasing observational precision from increasing observational breadth.

The same principle applies more generally to precision cosmology. As measurement uncertainties decrease, correlations arising from shared data, common analysis pipelines, repeated physical targets and widely reused benchmark values become proportionately more important. A literature may converge tightly for sound scientific reasons while nevertheless containing substantially fewer independent constraints than its publication count suggests. Provenance therefore becomes more, rather than less, important as a field enters the precision regime.

For primordial D/H, the encouraging result of the present analysis is that accounting for these dependencies does not undermine the modern central abundance. The convergence near $10^5({\rm D/H})\simeq2.5$ is recovered under several alternative treatments and persists beyond the small group of absorbers that dominates the published record. The challenge for future work is therefore not primarily to establish whether a consensus exists but to ensure that increasingly precise cosmological inferences properly reflect the number, diversity and statistical relationships of the physical systems on which that consensus rests.

\section{Conclusions}
\label{conclusions.sec}

We have reconstructed the development of published primordial deuterium abundances from 1994 to 2026 using a provenance-aware database containing 957 reported D/H values from 247 publications. Unlike conventional compilations designed primarily to estimate a preferred abundance, our analysis treats the published literature itself as an object of study. We distinguish newly generated individual measurements and compilation values from their subsequent
reuse, and trace observational entries wherever possible to the physical absorption systems from which they ultimately derive.

The resulting picture is strongly hierarchical. Of the 795 observational entries in our database, 80.1 percent represent reused individual or compilation values, whereas only 91 entries (11.4 percent) correspond to new determinations of individual absorption systems. Although 52 distinct physical absorbers can be identified, their representation is highly unequal: the eight most frequently represented systems account for 62.0 percent of all identifiable absorber mentions, and the inverse Herfindahl--Hirschman concentration measure corresponds to only $N_{\rm eff}=16.28$ equally represented absorbers. The numerical size of the literature therefore substantially exceeds the breadth of its underlying observational foundation.

This dependence does not invalidate the modern primordial-D/H consensus. Sensitivity tests using new individual measurements published from 2012 onwards show that treating every published value as independent can give substantial leverage to repeatedly analysed absorbers and formal uncertainties that do not reflect this absorber-level structure. Nevertheless, alternative treatments based on individual absorbers continue to recover abundances close to $10^5({\rm D/H})\simeq2.5$. The median of the 15 absorber-level medians is $10^5({\rm D/H})=2.5000$, whereas removal of all eight systems that dominate the publication-frequency distribution gives $10^5({\rm D/H})=2.4960\pm0.0265$ for the remaining modern measurements. The central abundance is therefore more robust than the genealogy of the literature might initially suggest, even though its effective statistical independence is considerably smaller than a simple count of published determinations would imply.

The historical analysis also shows that convergence was not simply the result of accumulating measurements. A conspicuous transition occurs in the compilation literature around 2013--2014 across all seven dominant absorbers with earlier publication histories, while reused individual values remain substantially more heterogeneous. This transition can be understood in the context of increasing concern over heterogeneous samples and unrecognised systematic uncertainties, followed by the development of deliberately restricted samples analysed within increasingly homogeneous frameworks. The
Precision Sample introduced by \citet{2014ApJ...781...31C} represents a particularly clear manifestation of this change. Its influence was subsequently amplified as the resulting compilation abundance became a recurring numerical reference point in later publications.

This distinction reveals two forms of inheritance within the literature. Individual absorbers can be revisited observationally, producing genuinely new information while retaining dependence on the same physical sightline. At the same time, previously published measurements and compilation values can acquire extended bibliographic lives through direct numerical reuse. Compilations can therefore function not only as summaries of existing measurements but also as nodes of numerical inheritance through which particular estimates propagate across subsequent generations of publications. The number of literature appearances, the number of measurements and the number of independent physical sightlines are consequently different quantities.

Our main conclusion is therefore twofold. The modern primordial-deuterium abundance is scientifically well supported: its central value remains stable under substantially different treatments of the observational genealogy and is not confined to the small set of absorbers that dominates the published record. Its apparent precision, however, requires more careful interpretation because shared absorbers, repeated analyses and inherited compilation values reduce the effective independence of the literature. Future progress will benefit not only from improved spectroscopy and reanalysis of established systems, but particularly from additional independent high-quality absorbers and from compilations that preserve explicit information about provenance and statistical relationships. As primordial D/H and other cosmological observables enter increasingly precise regimes, such provenance-aware treatment becomes more, rather than less, important.

\bibliographystyle{mnras}
\bibliography{deuterium} 


\appendix
\renewcommand{\thetable}{A\arabic{table}}
\setcounter{table}{0}

\onecolumn
\section{Unique absorber dataset}
\label{absorbers.sec}

Table \ref{tab:absorber_census} provides an overview of the absorption systems represented in the primordial-D/H master database. The cleaned database contains 52 identifiable physical absorption systems. Fifty are distinguished by sightline and absorber redshift; two additional named systems are retained although a secure absorber redshift could not be assigned from the source material.

The table is derived directly from the provenance database. Full source-level references for individual entries are available in the accompanying online database and are not repeated here. A small number of database entries refer to one or more absorption systems that could not be identified individually from the source publication; these are retained in the global literature analysis but are omitted from this absorber-level catalogue.

\begin{small}
\setlength{\tabcolsep}{3pt}
\begin{longtable}{
>{\raggedright\arraybackslash}p{2.6cm}
>{\raggedright\arraybackslash}p{2.0cm}
c
r
r
>{\raggedright\arraybackslash}p{4.2cm}
}
\caption{Catalogue of the identified absorption systems represented in the primordial-D/H database. For each system we list the absorber redshift, the year of its first D/H appearance in the database, the total number of database appearances ($N_{\rm app}$), and the number of entries classified as new individual measurements ($N_{\rm new}$). Full source-level provenance is provided in the accompanying online database.}
\label{tab:absorber_census}\\
\hline
Absorber (sightline) & $z_{\rm abs}$ & First D/H appearance & $N_{\rm app}$ & $N_{\rm new}$ & Notes \\
\hline
\endfirsthead
\multicolumn{6}{c}{\tablename\ \thetable\ --- continued from previous page}\\
\hline
Absorber (sightline) & $z_{\rm abs}$ & First D/H appearance & $N_{\rm app}$ & $N_{\rm new}$ & Notes \\
\hline
\endhead
\multicolumn{6}{r}{Continued on next page}\\
\endfoot
\bottomrule
\endlastfoot
APM08279+5255 & 3.514 & 1999 & 4 & 1 &  \\
BR1202--0725 & 4.383 & 1996 & 1 & 1 &  \\
BR1202--0725 & 4.672 & 1996 & 8 & 0 &  \\
CTQ247 & 2.621 & 2023 & 9 & 0 &  \\
HS0105+1619 & 2.53651 & 2000 & 212 & 2 &  \\
HS0105+1619 (2) & 2.53651 & 2016 & 8 & 0 & Secondary canonical designation retained from the master database. \\
J0407--4410 & 2.618 & 2012 & 37 & 1 &  \\
J0744+2059 & 2.5338836 & 2016 & 9 & 1 &  \\
J0812+3208 & 2.626 & 2010 & 4 & 2 &  \\
J1134+5742 & 3.05 & 2017 & 2 & 0 &  \\
J1134+5742 & 3.41088 & 2012 & 30 & 0 &  \\
J1134+5742 & 3.522 & 2011 & 30 & 1 &  \\
J1201+0116 & 2.98838 & 2016 & 9 & 1 &  \\
J1254+3627 & 2.985 & 1998 & 1 & 0 &  \\
J1332+0052 & 3.42 & 2023 & 15 & 2 &  \\
J1337+3152 & 3.102 & 2012 & 17 & 0 &  \\
J1337+3152 & 3.16768 & 2010 & 26 & 1 &  \\
J1358+0349 & 2.85305 & 2016 & 81 & 1 &  \\
J1358+6522 & 3.06726 & 2014 & 111 & 1 &  \\
J1419+0829 (2) & 3.04973 & 2014 & 78 & 0 & Secondary canonical designation retained from the master database. \\
J1419+0829 & 3.04984 & 2012 & 51 & 2 &  \\
J1439+1117 & 2.418 & 2016 & 1 & 0 &  \\
J1444+2919 & 2.437 & 2016 & 60 & 1 &  \\
J1558--0031 & 2.70242 & 2006 & 140 & 2 &  \\
J1558--0031 (2) & 2.70242 & 2016 & 8 & 0 & Secondary canonical designation retained from the master database. \\
J1558--0031 & 2.823 & 2012 & 8 & 0 &  \\
J1757+7539 & 2.62 & 1998 & 1 & 0 &  \\
J1757+7539 & 2.91 & 1998 & 1 & 0 &  \\
J2123--0050 & 2.059 & 2016 & 1 & 0 &  \\
PKS1937--101 (2) & 3.25601 & 2004 & 73 & 3 & Secondary canonical designation retained from the master database. \\
PKS1937--101 & 3.5722 & 1994 & 258 & 23 &  \\
Q0014+813 & 2.797957 & 1996 & 5 & 2 &  \\
Q0014+813 & 3.320482 & 1994 & 70 & 8 &  \\
Q0130--403 & 2.799 & 1999 & 33 & 2 &  \\
Q0347--383 & 3.025 & 2001 & 54 & 3 &  \\
Q0420--388 & 3.086 & 1996 & 13 & 2 &  \\
Q0454--220 & 0.482 & 1996 & 2 & 0 &  \\
Q0454--220 & 0.535 & 1999 & 1 & 0 &  \\
Q0636+680 & 2.8904 & 1996 & 2 & 0 &  \\
Q0913+072 & 2.61843 & 2008 & 137 & 3 &  \\
Q0913+072 & 2.785 & 2008 & 9 & 0 &  \\
Q1009+2956 & 2.50357 & 1996 & 231 & 13 &  \\
Q1232+082 & 2.3377 & 2010 & 4 & 1 &  \\
Q1243+307 & 2.426 & 2017 & 7 & 0 &  \\
Q1243+307 & 2.52566 & 2003 & 149 & 3 &  \\
Q1251+3644 & \ldots & 1997 & 1 & 0 & No absorber redshift is assigned in the master database. \\
Q1331+170 & 1.777 & 2010 & 4 & 2 &  \\
Q1422+230 & 3.478 & 1998 & 2 & 1 &  \\
Q1422+230 & 4.515 & 1997 & 1 & 0 &  \\
Q1718+4807 & 0.701 & 1997 & 43 & 5 &  \\
Q1759+7539 & \ldots & 1997 & 1 & 0 & No absorber redshift is assigned in the master database. \\
Q2206--199 & 2.07624 & 2001 & 110 & 1 &  \\
\hline
\end{longtable}
\end{small}

\end{document}